\documentclass{article}
\usepackage{subcaption}
\usepackage{arxiv}
\usepackage{array}
\usepackage{amsmath}
\usepackage{tikz}
\usepackage[utf8]{inputenc} 
\usepackage[T1]{fontenc}    
\usepackage{hyperref}       
\usepackage{url}            
\usepackage{booktabs}       
\usepackage{amsfonts}       
\usepackage{nicefrac}       
\usepackage{microtype}      
\usepackage{lipsum}
\usepackage{graphicx}
\usepackage{amsmath}
\usepackage{amsfonts}
\usepackage{algorithm}
\usepackage{algorithmic}
\usepackage[numbers]{natbib}
\usepackage{booktabs}
\usepackage{multirow}

\graphicspath{ {./images/} }

\title{Sig2text, a Vision-language model for Non-cooperative Radar Signal Parsing}

\author{
 Hancong Feng \\
 School of Information and Communication Engineering\\
 UEST of China\\
  \texttt{202211012303@std.uestc.edu.cn} \\
   \And
 KaiLi Jiang \\
 School of Information and Communication Engineering\\
 UEST of China\\
  \texttt{jiangkelly@foxmail.com} \\
  \And
 Bin tang \\
 School of Information and Communication Engineering\\
 UEST of China\\
}

\begin{document}
\sloppy
\maketitle
\begin{abstract} 
Automatic non-cooperative analysis of intercepted radar signals is essential for intelligent equipment in both military and civilian domains. Accurate modulation identification and parameter estimation enable effective signal classification, threat assessment, and the development of countermeasures. In this paper, we propose a symbolic approach for radar signal recognition and parameter estimation based on a vision-language model that combines context-free grammar with time-frequency representation of radar waveforms. The proposed model, called Sig2text, leverages the power of vision transformers for time-frequency feature extraction and transformer-based decoders for symbolic parsing of radar waveforms. By treating radar signal recognition as a parsing problem, Sig2text can recognize the types and estimating the parameters of signals with arbitrarily complex modulations. We evaluate the performance of Sig2text on a synthetic radar signal dataset and demonstrate its effectiveness in various scenarios.

\end{abstract}

\section{Introduction}
Automatic non-cooperative analysis of intercepted radar signals plays a crucial role in intelligent systems for both military \cite{zohuri2020radar} and civilian applications \cite{ma2020joint,aydogdu2020radar}. Accurate modulation identification and parameter estimation facilitate effective signal classification, threat assessment, and countermeasure development. However, the widespread adoption of software-defined radar systems \cite{gurbuzOverviewCognitiveRadar2019, bluntOverviewRadarWaveform2016} has introduced new challenges. Modern radar systems employ advanced techniques such as low-power transmission, spread spectrum methods, frequency agility, and complex modulation schemes, significantly complicating signal analysis \cite{paceDetectingClassifyingLow2009, taoResearchLPIRadar2021}.

Traditional approaches for radar signal intra-pulse modulation recognition and parameter estimation primarily rely on statistical signal processing techniques. Modulation recognition typically involves feature extraction followed by classification using statistical pattern recognition algorithms. For instance, \cite{lundenAutomaticRadarWaveform2007} extracted features from Wigner and Choi-Williams time-frequency distributions and pruned redundant ones using an information-theoretic approach. Similarly, \cite{zhuClassificationRadarEmitter2007} decomposed signals into time-frequency atoms via a fast matching pursuit algorithm for robust emitter signature clustering. Decision-theoretic frameworks and statistical tests were introduced for intrapulse modulation recognition in \cite{mingIntrapulseModulationRecognition2011,zengApproachIntraPulseModulation2010}. Parameter estimation methods have addressed challenges posed by low signal-to-noise ratios and complex modern radar signals. \cite{chilukuriEstimationModulationParameters2020} employed a cyclostationary approach to extract key modulation parameters of low probability of intercept (LPI) radar signals, achieving estimation errors below 6\%. \cite{baiChirpRateEstimation2019} reformulated chirp rate estimation as frequency estimation problems using multiple discrete polynomial phase transforms with optimal weighting. Additionally, \cite{taoResearchLPIRadar2021} integrated time-frequency representations with visibility graph techniques for multicomponent LPI signal detection. Despite their effectiveness, these methods often depend on handcrafted features for specific signal types and struggle to adapt to emerging complex modulation schemes.

The rapid advancement of deep learning has led to numerous approaches, broadly categorized by their input: direct sampled sequences from analog-to-digital converters (ADCs) and time-frequency image representations. Sequence-based methods process sampled radar signals directly through neural networks for automatic feature extraction and modulation classification. Studies employing convolutional neural networks (CNNs) and long short-term memory (LSTM) networks have demonstrated promising results \cite{jingAdaptiveFocalLoss2022,weiIntrapulseModulationRadar2020,sunRadarEmitterClassification2018,yuanIntraPulseModulationClassification2021}. Time-frequency image (TFI)-based approaches transform radar signals into time-frequency images before processing them with image-based neural networks. Specialized architectures \cite{huynh-theAccurateLPIRadar2021,quRadarSignalIntrapulse2019,guoLPIRadarWaveform2019,guoRadarSignalRecognition2024} have shown significant potential. For more complex scenarios involving overlapping signals or hybrid modulations, multi-label classification methods have been explored, including residual networks \cite{hong-haiRadarEmitterMultilabel2022}, multi-instance learning \cite{pan2020automatic}, feature pyramid networks with class activation mapping \cite{siMultilabelHybridRadar2022}, and deep multi-label classifiers \cite{zhu2020automatic}.

Existing methods primarily focus on intra-pulse modulation recognition without performing parameter estimation. To address this limitation, multi-task learning approaches have emerged. \cite{akyonMultiTaskLearningBased2019} proposed a multi-task learning technique using recurrent neural networks for joint SNR estimation, pulse detection, and modulation classification. \cite{zhuJMRPENetJointModulation2021} introduced JMRPE-Net, a deep multitask network for joint modulation recognition and parameter estimation of cognitive radar signals. For overlapped signals, \cite{chenSemanticLearningAnalysis2023} developed a joint semantic learning deep convolutional neural network that simultaneously performs feature restoration, modulation classification, and parameter regression. However, these models typically have fixed output heads, limiting their adaptability to signals with varying parameter counts and complex hybrid modulations.

This paper presents a symbolic approach for radar signal recognition and parameter estimation based on a vision-language model that combines context-free grammar (CFG) with time-frequency representations. Our proposed model, Sig2text, leverages vision transformers for time-frequency feature extraction and transformer-based decoders for symbolic waveform parsing. By framing radar signal recognition as a parsing problem, Sig2text effectively recognizes and parses waveforms with diverse modulation types and parameters. We evaluate Sig2text on a synthetic radar signal dataset, demonstrating its effectiveness in both recognition and parameter estimation.

The main contributions of this paper are:
\begin{itemize}
    \item A symbolic radar signal representation method that compactly encodes arbitrary waveforms with varying parameters
    \item Sig2text, a novel architecture combining vision transformers for feature extraction with transformer-based decoders for symbolic parsing
    \item Comprehensive evaluation on synthetic radar signals, demonstrating effective type recognition and parameter estimation
\end{itemize}

\section{Problem Formulation}
\subsection{Signal Model}\label{sec:signal_model}
For a non-cooperative radar signal receiver, the sampled signal $y(k)$ with additive white Gaussian noise (AWGN) can be modeled as:
\begin{align}
    \label{signal_model}
    y(k)=A\exp \left(j \cdot\left(\theta(k)+2 \pi \frac{f_{c}}{f_s} k\right)+\theta_{p}\right)+n(k)
\end{align}
where $A$ denotes the constant signal amplitude within the pulse width, $\theta(k)$ represents the instantaneous phase, $f_c$ is the carrier frequency, $f_s$ is the sampling frequency, $\theta_p$ is the phase offset, and $n(k)$ is the AWGN component.

Based on the instantaneous phase $\theta(k)$, fundamental radar signals can be categorized into three primary types: frequency-modulated (FM), phase-coded (PC), and frequency-coded (FC) waveforms \cite{bluntOverviewRadarWaveform2016}. FM waveforms employ continuous phase modulation for pulse compression, with linear frequency modulation (LFM) being the most prevalent example. PC waveforms divide the pulse into constant-amplitude chips, each modulated by discrete phase values (either binary or polyphase). FC waveforms modulate their frequency according to predetermined stepped-frequency sequences, with Costas codes representing a common implementation. Table \ref{tab:basic_waveforms} summarizes these fundamental radar waveforms and their principal subclasses.

\begin{table}[h]
  \centering
  \renewcommand{\arraystretch}{1.3}
  \caption{Basic radar waveforms and their subclasses}
  \begin{tabular}{l|l}
      \toprule
      \textbf{Waveform Type} & \textbf{Subclasses} \\
      \midrule
      FM & LFM, Triangular Waveform (Tri) \\
      PC & Barker, Polyphase, Frank, P1, P2, P3, P4 \\
      FC & Costas \\
      \bottomrule
  \end{tabular}
  \label{tab:basic_waveforms}
\end{table}

Beyond these fundamental waveforms, practical radar systems often employ hybrid modulation schemes. For instance, combining FM and PC waveforms can produce signals like LFM/binary phase-shift keying (BPSK) to achieve low probability of interception. Similarly, integrating FC and PC waveforms enables effective jamming suppression \cite{feiHybridFSKPSKWaveform2021}. Multiple LFM components with different slopes can also be combined to enhance velocity estimation in synthetic aperture radar (SAR) systems \cite{zhaoNovelApproachSlope2018}.

Given the exponential growth of possible signal combinations and their uncertain quantities, radar signal analysis cannot be adequately addressed as a simple classification or regression problem. We therefore formulate it as a parsing problem from the signal domain to the symbolic domain, as detailed in the following section.

\subsection{The Task of Radar Signal Parsing}
Similar to how humans use natural language with finite alphabets to describe image content, we can define a formal language to describe radar signals. Let $\Sigma$ be a finite alphabet (a finite set of symbols), and let $\Sigma^*$ denote the set of all finite strings over $\Sigma$:
\[
\Sigma^* = \bigcup_{n=0}^\infty \Sigma^n,
\]
where $\Sigma^n$ represents all strings of length $n$ (with $\Sigma^0 = \{\varepsilon\}$, where $\varepsilon$ is the empty string). A \emph{formal language} is any subset of $\Sigma^*$:
\[
L \subseteq \Sigma^*.
\]

Although the alphabet $\Sigma$ is finite, $\Sigma^*$ is infinite due to arbitrary symbol concatenation. This property provides formal languages with substantial expressive power, enabling the encoding of unlimited information. Furthermore, many formal languages employ grammars with recursive production rules, facilitating the construction of nested hierarchical structures necessary for representing complex syntactic patterns.

We formalize radar signal parsing as a mapping from continuous signals to discrete symbolic representations. Let a radar signal be represented by a vector
\[
I \in \mathbb{R}^{N \times 1},
\]
where $N$ is the number of sample points. Given a finite alphabet $\Sigma$ and the set of all finite strings $\Sigma^*$, the symbolic description of a radar signal is a sequence
\[
T = (w_1, w_2, \dots, w_n) \in \Sigma^*,
\]
where each $w_t \in \Sigma$. Our objective is to learn a mapping
\[
f: \mathbb{R}^{N \times 1} \rightarrow \Sigma^*
\]
by maximizing the conditional probability $P(T \mid I)$. Applying the chain rule yields the factorization:
\[
P(T \mid I) = \prod_{t=1}^{n} P\bigl(w_t \mid I, w_1, \dots, w_{t-1}\bigr).
\]
Thus, the parsing problem reduces to a conditional next-token prediction task, where learning $f$ corresponds to estimating $P(w_t \mid I, w_1, \dots, w_{t-1})$ through neural network training.

The language design critically affects the clarity and conciseness of radar signal representation. In the following section, we present a specially designed language capable of representing arbitrary radar waveforms with varying parameters.

\section{Method}
This section first presents the symbolic representation of radar waveforms, then introduces the vision transformer architecture for time-frequency feature extraction and transformer-based decoders for symbolic parsing.

\subsection{Symbolic Representation of Radar Waveforms}
We describe an artificial language for representing arbitrary radar waveforms with varying parameters, enabling the formulation of radar signal recognition and parameter estimation as a language parsing problem.

The primary consideration is language type selection. While natural language descriptions are possible, they prove inefficient due to their extensive vocabularies (e.g., English or Chinese) and poor machine-parsability. Given the limited number of radar signal types and parameters, we employ a context-free grammar (CFG) for language design.

A CFG is a formal grammar where each production rule takes the form $A \rightarrow \alpha$, with $A$ as a non-terminal symbol and $\alpha$ as a string of terminal and/or non-terminal symbols. The language generated by a CFG comprises all strings derivable from the start symbol through production rule applications. For instance, the CFG $G = (\{S\}, \{a, b\}, \{S \rightarrow aSb, S \rightarrow \epsilon\}, S)$ generates the language $\{a^nb^n | n \geq 0\}$. Despite their simple rules, CFGs can generate complex languages, as demonstrated in image processing \cite{chua2021deepcpcfg} and multifunction radar modeling \cite{visnevskiSyntacticModelingSignal2007}.

For basic waveforms (FM, phase-coded, and frequency-coded), our string representations follow the pattern: \texttt{<type> <sub> <parameter\_list>}. The \texttt{type} field indicates the modulation category ('FM', 'PC', or 'FC'), while \texttt{sub\_type} specifies the waveform variant (e.g., 'barker' under 'PC'). The \texttt{parameter\_list} contains signal parameters like bandwidth, carrier frequency, and code sequences. Table \ref{tab:simple_grammar_rules} presents the grammar rules. For example:
\begin{itemize}
    \item \texttt{FM LFM cf 100.0 B 10.0} denotes a linear FM waveform with 100.0 MHz center frequency and 10.0 MHz bandwidth
    \item \texttt{FC Costas cf 100.0 FH 1.0 Code 7 6 2 10 1 4 8 9 11 5 3} represents a Costas-coded waveform with 100.0 MHz center frequency, 1.0 MHz frequency hop, and specified code sequence
\end{itemize}

\begin{table}[h]
  \centering
  \renewcommand{\arraystretch}{1.3}
  \caption{Grammar rules for five basic radar waveforms (LFM, P1, P2, P4, CostasFM). Terminal symbols: $<FM>$, $<PM>$, $<FC>$, $<P1>$, $<P2>$, $<P4>$, $<Costas>$, $<LFM>$ (waveform types); $<cf>$, $<B>$, $<FH>$, $<Code>$ (estimated parameters); $<FLOAT>$, $<INT>$ (parameter values).}
  \begin{tabular}{l|l}
      \toprule
      \textbf{Non-terminal} & \textbf{Productions} \\
      \midrule
      $<ENTRY>$ & $<FM\_ENTRY> \mid <PM\_ENTRY> \mid <FC\_ENTRY>$ \\
      $<FM\_ENTRY>$ & $<FM> <LFM> <cf> <FLOAT> <B> <FLOAT>$ \\
      $<PM\_ENTRY>$ & $<PM> <P\_TYPE> <cf> <FLOAT> <code\_length> <INT>$ \\
      $<FC\_ENTRY>$ & $<FC> <Costas> <cf> <FLOAT> <FH> <FLOAT> <Code> <INT\_LIST>$ \\
      $<P\_TYPE>$ & $<P1> \mid <P2> \mid <P4>$ \\
      $<INT\_LIST>$ & $<INT> <INT\_LIST> \mid <INT>$ \\
      \bottomrule
  \end{tabular}
  \label{tab:simple_grammar_rules}
\end{table}

Composite waveforms are handled through recursive grammar rules by adding $<S>\rightarrow <ENTRY> <S> \mid <ENTRY>$. For example, \texttt{FM LFM cf 100.0 B 10.0 PM P1 cf 100.0 code\_length 10} combines LFM and P1 waveforms with repeated carrier frequency (only the first occurrence is retained during parsing).

Continuous parameter values (e.g., $<FLOAT>$) require quantization into discrete tokens for decoder compatibility. The quantization process involves dividing by an appropriate unit and rounding to the nearest integer. For instance, with a 0.1 MHz quantization unit, a 23.43 MHz bandwidth becomes token 234.

\subsection{Architecture of Sig2text}

To enhance feature interpretability and noise robustness, we first apply time-frequency transformation to radar waveforms before encoder processing. For computational efficiency, we employ the Short-Time Fourier Transform (STFT), retaining only the magnitude component which contains the most salient information, resulting in single-channel time-frequency representations.

While convolutional neural networks (CNNs) have traditionally dominated time-frequency image processing, their limited receptive fields constrain their ability to capture long-range dependencies. To overcome this limitation, we adopt the Vision Transformer (ViT) \cite{dosovitskiy2020image}, which processes images as sequences of patches through transformer encoders. This architecture enables global receptive fields through self-attention mechanisms, demonstrating superior performance in various vision tasks including recognition and segmentation \cite{zhangVisionLanguageModelsVision2024}.

For a time-frequency image of dimensions $N \times M$, we partition it into $n \times m$ patches, where $n$ divides $N$ and $m$ divides $M$. This flexible patching scheme accommodates varying pulse widths in radar waveforms. Each patch is flattened and linearly projected into a $D$-dimensional embedding space, producing input vectors $\mathbb{R}^{\left(\frac{N}{n} \times \frac{M}{m}\right) \times D}$ for the transformer.

To preserve spatial information, we augment patch embeddings with learnable positional encodings $P \in \mathbb{R}^{L \times D}$, where $L=\left(\frac{N}{n} \times \frac{M}{m}\right)$ represents the total patch count. These trainable embeddings enable the model to learn optimal spatial representations during training.

The encoder comprises $L$ identical layers, each featuring:
\begin{itemize}
    \item Multi-head self-attention mechanisms
    \item Layer normalization
\end{itemize}
Following \cite{vaswaniAttentionAllYou2017}, the final encoder layer projects outputs to a $K$-dimensional vector for decoder processing.

\begin{figure}[H]
  \centering
  \includegraphics[width=1\textwidth]{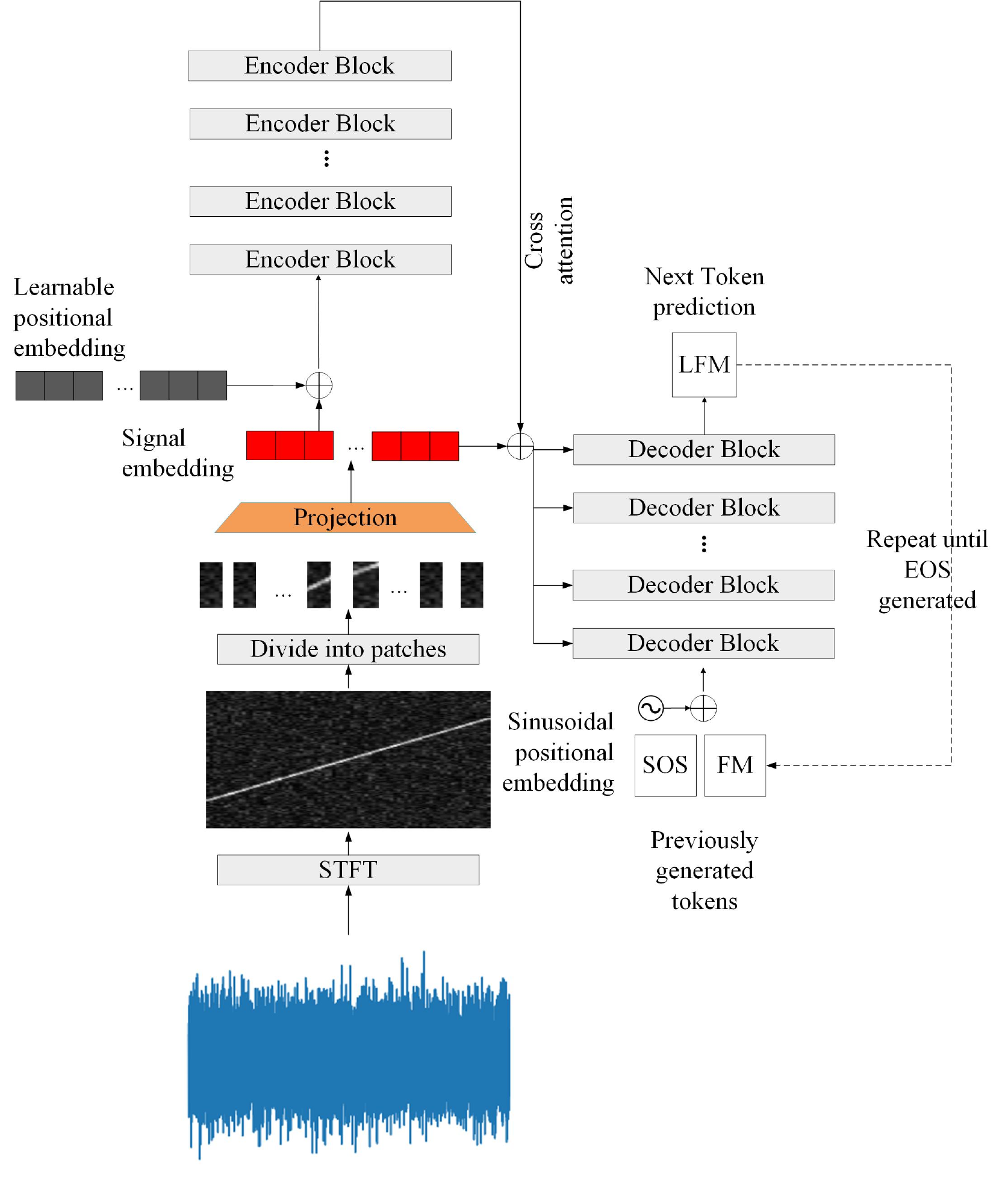}
  \caption{Architecture overview of Sig2text, demonstrating the complete processing pipeline from raw signal input to symbolic output.}
  \label{fig:network_architecture}
\end{figure}

The decoder similarly consists of $L$ identical layers with:
\begin{itemize}
    \item Masked multi-head self-attention
    \item Layer normalization
    \item Residual connections between embedding and decoder layers
\end{itemize}
The decoder outputs are linearly transformed to $K$-dimensional predictions for next-token generation. Figure \ref{fig:network_architecture} illustrates the complete Sig2text architecture, detailing the signal-to-symbol transformation pipeline.

\subsection{Training and Inference}
Given a dataset of radar signals $\{x^{(i)}\}_{i=1}^N$ with corresponding symbolic representations $\{s^{(i)}\}_{i=1}^N$, our objective is to learn model parameters $\theta$ that maximize the conditional likelihood:

\begin{equation}
    \mathcal{L}(\theta) = \frac{1}{N}\sum_{i=1}^{N}\log p_{\theta}(s^{(i)}|x^{(i)})
\end{equation}

Direct likelihood maximization through autoregressive prediction can be computationally inefficient. We therefore employ teacher forcing during training, using ground truth tokens as decoder inputs rather than model predictions. This yields the training objective:

\begin{equation}
    \mathcal{L}_{\text{TF}}(\theta) = -\frac{1}{N}\sum_{i=1}^{N}\sum_{t=1}^{T_{i}}\log p_{\theta}(s_t^{(i)}|s_{<t}^{(i)},x^{(i)})
\end{equation}

where $T_i$ denotes the length of $s^{(i)}$, and $p_{\theta}(s_t^{(i)}|s_{<t}^{(i)},x^{(i)})$ represents the conditional probability of token $s_t^{(i)}$ given previous tokens $s_{<t}^{(i)} = (s_1^{(i)},...,s_{t-1}^{(i)})$ and input $x^{(i)}$. We optimize this objective via gradient descent, with the complete training procedure outlined in Algorithm~\ref{alg:training}. Special tokens $<sos>$ and $<eos>$ mark the start and end of each string respectively.

\begin{algorithm}[H]
\caption{Model Training Procedure}
\label{alg:training}
\begin{algorithmic}[1]
\REQUIRE Training set $\mathcal{D} = \{(x^{(i)}, s^{(i)})\}_{i=1}^N$, learning rate $\eta$, epochs $E$, batch size $B$
\STATE Quantize continuous parameters in $s^{(i)}$ to discrete tokens
\STATE Augment strings with $<sos>$ and $<eos>$ tokens; construct vocabulary
\FOR{$epoch = 1$ \TO $E$}
  \FOR{each mini-batch $\mathcal{B} \subset \mathcal{D}$}
    \STATE Compute STFT magnitude spectra for signals $x$
    \STATE Partition time-frequency images into patches; project to $D$-dim embeddings
    \STATE Incorporate learnable positional encodings
    \STATE Encode patches via Transformer encoder
    \STATE Predict next-token distributions using Transformer decoder
    \STATE Compute negative log-likelihood loss (equivalent to cross-entropy)
    \STATE Update parameters via gradient descent with learning rate $\eta$
  \ENDFOR
  \STATE \textbf{Optional:} Evaluate on validation set
\ENDFOR
\end{algorithmic}
\end{algorithm}

During inference, we employ beam search decoding with width $K$ to generate predictions. The algorithm maintains $K$ most probable partial sequences at each step, expanding and pruning candidates until termination conditions are met. The complete inference process is detailed in Algorithm~\ref{alg:inference}.

\begin{algorithm}[H]
\caption{Beam Search Inference} 
\label{alg:inference}
\begin{algorithmic}[1]
\REQUIRE Input signal $x$, beam width $K$, maximum length $T_{\text{max}}$
\STATE Compute STFT magnitude spectrum of $x$
\STATE Process time-frequency image through patch embedding layer
\STATE Add positional encodings
\STATE Generate latent representation via Transformer encoder
\STATE Initialize beam with $<sos>$ token
\FOR{$t = 1$ \TO $T_{\text{max}}$}
  \FOR{each partial sequence in beam}
    \STATE Predict next-token distribution using decoder
    \STATE Compute sequence log-likelihoods
    \STATE Expand partial sequences with candidate tokens
  \ENDFOR
  \STATE Retain top-$K$ sequences by likelihood
  \IF{any sequence terminates with $<eos>$}
    \STATE Break
  \ENDIF
\ENDFOR
\STATE Return highest-likelihood sequence, converting tokens to waveform parameters
\end{algorithmic} 
\end{algorithm}
\section{Simulations}
This section evaluates the performance of our proposed Sig2text model on synthetic radar signal datasets. We assess the model's capabilities in both type recognition and parameter estimation, while also examining its robustness and transfer learning potential for real-world applications. The PyTorch implementation is publicly available\textsuperscript{\href{https://github.com/Na-choneko/sig2text}{1}}.

\subsection{Dataset Description}
We evaluate our model using synthetic datasets generated by a CUDA-based radar simulator that models the complete transmission and reception chain. The ADC the receiver is 100 MHz with 8-bits effective precision. The pulse widths of the emitter ranging from 50 $\mu$s to 100 $\mu$s. Due to different Effective Radiated Power(ERP) and distance from the emitter to the receiver, the signal-to-noise ratio (SNR) of the signals varies from -10dB to 10dB.

Our first dataset contains 500,000 samples uniformly distributed across 13 classes:
\begin{itemize}
    \item Frequency modulation: LFM, sinusoidal (Sin), triangular (Tri)
    \item Phase coding: Frank, P1-P4, T1-T4
    \item Frequency coding: Costas
\end{itemize}

Table \ref{tab:modulation_parameters} details the parameter ranges for each modulation type. All parameters were randomly generated within these specified bounds.

\begin{table}[htbp] 
  \centering
  \caption{Parameter Ranges for LPI Radar Signals}
  \label{tab:modulation_parameters}
  \begin{tabular}{llll}
  \toprule
  \textbf{Parameter} & \textbf{Range} & \textbf{Applicable Signals} & \textbf{Description} \\
  \midrule
  $cf$ & 10--40 MHz & All & Center frequency \\
  $B$ & 2--20 MHz & LFM: 2--10; Sin/Tri: 10--20 & Bandwidth \\
  $T$ & 5--20 $\mu$s & Sin/Tri & Modulation period \\
  $FH$ & $\geq$0.5 MHz & Costas & Frequency hop step \\
  segment\_num & 1--4 & T1-T4 & Segment count \\
  phasestate\_num & 2--4 & T1-T4 & Phase states \\
  $\Delta F$ & 1--10 MHz & T3/T4 & Modulation bandwidth \\
  code\_length & 3--10 & P1-P4 & Code length \\
  Code & - & Costas & Costas sequences \\
  \bottomrule
  \end{tabular}
\end{table}

The second dataset extends the first by adding 500,000 hybrid modulation samples. These combine fundamental waveforms through various composition methods, as detailed in Table \ref{tab:hybrid_modulations}.

\begin{table}[ht]
  \centering
  \caption{Hybrid Modulation Combinations}
  \label{tab:hybrid_modulations}
  \begin{tabular}{lll}
  \toprule
  \textbf{Hybrid Type} & \textbf{Components} & \textbf{Combination Method} \\
  \midrule
  FM+FM & Multiple LFM & Sequential \\
  FM+PM & LFM + Polyphase (P1-4, T1-4, Frank) & Multiplicative \\
  FM+FC & LFM + Costas & Multiplicative \\
  PM+FC & Polyphase + Costas & Multiplicative \\
  PM+PM & Two Polyphase (P1-4, Frank) & Sequential \\
  \bottomrule
  \end{tabular}
\end{table}

\subsection{Hyperparameters and Baseline Methods}
This section details the experimental hyperparameters and baseline methods used for comparative evaluation.

Since existing baselines cannot perform complete signal recognition and parameter estimation for hybrid modulations, we compare full functionality only for basic modulations. For hybrid modulations, we evaluate only recognition performance. The baseline methods include:

\begin{itemize}
  \item \textbf{JMRPE-Net \cite{zhuJMRPENetJointModulation2021}}: A multi-task network combining CNN, LSTM, and self-attention modules for joint modulation recognition and parameter estimation.
  \item \textbf{MTViT}: A vision transformer encoder with fixed output heads, used to validate the feature extraction capability of our encoder architecture.
  \item \textbf{ResNet-based Multi-Label Classifier \cite{hong-haiRadarEmitterMultilabel2022}}: A CNN model employing residual blocks for time-frequency feature extraction, trained with multi-label classification loss.
\end{itemize}

Table \ref{tab:hyperparameters} summarizes the key hyperparameters for all models. We employ smaller Sig2text variants for single modulations and larger ones for hybrid modulations.

\begin{table}[h] 
  \centering
  \caption{Model Architecture Hyperparameters}
  \label{tab:hyperparameters}
  \begin{tabular}{|l|c|c|c|c|}
  \hline
  \textbf{Component} & \textbf{JMRPE-Net} & \textbf{MTViT} & \textbf{Sig2text-S} & \textbf{Sig2text-L} \\
  \hline
  CNN Layers & \begin{tabular}{@{}c@{}} 3 layers \\ (1,64,7) → (64,128,5) → \\ (128,256,3) \end{tabular} & - & - & - \\
  \hline
  LSTM & \begin{tabular}{@{}c@{}} Bidirectional \\ Input: 256 \\ Hidden: 128 \end{tabular} & - & - & - \\
  \hline
  Attention & \begin{tabular}{@{}c@{}} 8-head \\ Hidden: 256 \end{tabular} & \begin{tabular}{@{}c@{}} 8-head \\ Hidden: 128 \end{tabular} & \begin{tabular}{@{}c@{}} 8-head \\ Hidden: 128 \end{tabular} & \begin{tabular}{@{}c@{}} 8-head \\ Hidden: 512 \end{tabular} \\
  \hline
  Transformer Encoder & - & \begin{tabular}{@{}c@{}} 3 layers \\ Hidden: 128 \\ FF: 512 \end{tabular} & \begin{tabular}{@{}c@{}} 3 layers \\ Hidden: 128 \\ FF: 512 \end{tabular} & \begin{tabular}{@{}c@{}} 6 layers \\ Hidden: 512 \\ FF: 1024 \end{tabular} \\
  \hline
  Transformer Decoder & - & - & \begin{tabular}{@{}c@{}} 3 layers \\ Hidden: 128 \\ FF: 512 \end{tabular} & \begin{tabular}{@{}c@{}} 6 layers \\ Hidden: 512 \\ FF: 1024 \end{tabular} \\
  \hline
  \end{tabular}
\end{table}

For Sig2text, we implement the CFG rules shown in Table \ref{tab:grammar_rules} to generate symbolic signal descriptions. Continuous values are quantized with 0.01 resolution. During inference, we use beam search with size 1 and maximum output length 50. To convert the produced string into types and parameters, the bottom-up parsing \cite{hopcroft2001introduction} algorithm is employed.

\begin{table}[ht]
  \centering
  \caption{CFG Rules for Radar Signal Description}
  \label{tab:grammar_rules}
  \begin{tabular}{ll}
  \toprule
  \textbf{Non-Terminal} & \textbf{Production Rule} \\
  \midrule
  $S$ & $SimpleString \mid S\ SimpleString$ \\
  $SimpleString$ & $TypeChoice\ SubTypeChoice\ para$ \\
  $TypeChoice$ & \texttt{"FM"} $\mid$ \texttt{"PM"} $\mid$ \texttt{"FC"} \\
  $SubTypeChoice$ & \texttt{"LFM"} $\mid$ \texttt{"Sin"} $\mid$ \texttt{"Tri"} $\mid$ \texttt{"Costas"} $\mid$ \texttt{"T1--T4"} $\mid$ \texttt{"P1--P4"} \\
  $para$ & $LFMpara \mid SinTripara \mid Costaspara \mid T12para \mid T34para \mid Ppara$ \\
  $LFMpara$ & \texttt{"cf"} $Cf$ \texttt{"B"} $Bv$ \\
  $SinTripara$ & \texttt{"cf"} $Cf$ \texttt{"B"} $Bv$ \texttt{"T"} $Tv$ \\
  $Costaspara$ & \texttt{"cf"} $Cf$ \texttt{"FH"} $FHv$ \texttt{"Code"} $CS$ \\
  $T12para$ & \texttt{"cf"} $Cf$ \texttt{"seg\_num"} $SN$ \texttt{"phasestate\_num"} $PSN$ \\
  $T34para$ & \texttt{"cf"} $Cf$ \texttt{"seg\_num"} $SN$ \texttt{"phasestate\_num"} $PSN$ \texttt{"deltaF"} $DF$ \\
  $Ppara$ & \texttt{"cf"} $Cf$ \texttt{"code\_length"} $CL$ \\
  $Cf$ & $Number$ \\
  $Bv$ & $Number$ \\
  $Tv$ & $Number$ \\
  $FHv$ & $Number$ \\
  $CS$ & $Number \mid Number\ \texttt{" "}\ CS$ \\
  $SN$ & $Number$ \\
  $PSN$ & $Number$ \\
  $DF$ & $Number$ \\
  $CL$ & $Number$ \\
  $SCL$ & $Number$ \\
  \bottomrule
  \end{tabular}
\end{table}

All models were trained using AdamW optimizer \cite{loshchilovdecoupled} with batch size 128 and initial learning rate 0.001, reduced by 0.5 every 20 epochs. We reserved 3,000 samples for validation, stopping training when validation loss plateaued for 10 epochs. The hardware used for training included a single NVIDIA A100 GPU with 40GB memory.

\subsection{Results and Discussion}
The models are evaluated on the test set using two key metrics: (1) the accuracy of type/code recognition and (2) the mean square error (MSE) of parameter estimation. These metrics are defined as follows:
\begin{equation}
    \text{Accuracy} = \frac{\text{Number of correct predictions}}{\text{Total number of predictions}}
\end{equation}
\begin{equation}
    \text{MSE} = \frac{1}{N}\sum_{i=1}^{N}(\hat{para}_i - para_i)^2
\end{equation}
where $\hat{para}_i$ denotes the predicted parameter value and $para_i$ represents the ground truth. A prediction is considered correct only if all possible types/codes in a signal are accurately identified. Incorrect predictions are excluded from MSE calculations.

We first evaluate the models using the radar simulator, generating 1000 samples of 13 basic radar waveforms with uniform distribution at 0 dB SNR. The results, presented in Table~\ref{tab:results_0db}, demonstrate that Sig2text consistently outperforms both JMRPENet and MTvit across most parameters, achieving significantly lower MSE values—particularly for frequency-domain parameters ($cf$ and $B$). However, MTvit exhibits superior performance in estimating the number of segments in T codes, where both Sig2text and JMRPENet show limitations. This performance discrepancy may arise from two key factors: (1) the difference in loss functions (Sig2text employs token-prediction-based cross-entropy loss, while MTvit uses regression loss) and (2) the distinct input features, with MTvit leveraging STFT results that demonstrate stronger noise robustness compared to JMRPENet.

For type recognition accuracy, the STFT-based approach achieves near-perfect classification performance (0.999), whereas JMRPENet trails with 0.944 accuracy. This result confirms that time-frequency image (TFI) processing maintains an advantage over raw sequence processing for modulation classification tasks.

\begin{table}[ht]
    \caption{Performance comparison of different methods at SNR = 0 dB}
    \centering
    \label{tab:results_0db}
    \begin{tabular}{lcccc}
    \toprule
    Metric & Sig2text & JMRPENet & MTvit \\
    \midrule
    cf            & 0.0003 & 0.0203 & 0.6852 \\
    B              & 0.0053 & 0.0222 & 0.1441 \\
    T             & 0.0034 & 0.0031 & 0.0980 \\
    segment\_num  & 1.4759 & 1.7730 & 0.1683 \\
    phasestate\_num & 0.0214 & 0.1639 & 0.1239 \\
    FH            & 0.0001 & 0.0093 & 0.0217 \\
    code\_length  & 0.0000 & 0.0358 & 0.0250 \\
    deltaF        & 0.0098 & 0.0010 & 0.0024 \\
    Type Accuracy     & 0.999 & 0.944 & 0.996 \\
    \bottomrule
    \end{tabular}
\end{table}

\begin{figure}[htbp]
    \centering
    \begin{subfigure}[b]{0.48\textwidth}
        \includegraphics[width=\textwidth]{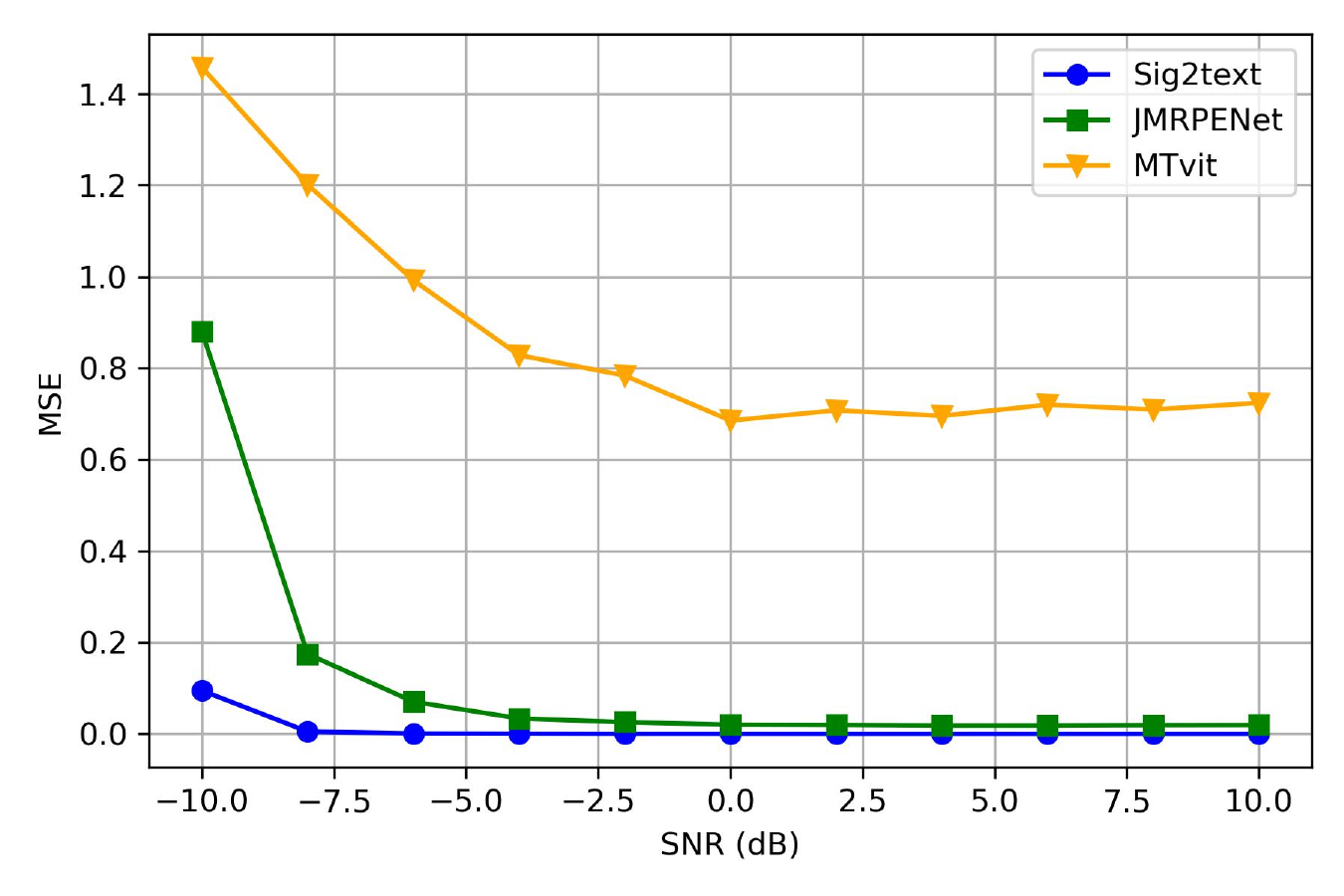}
        \caption{cf}
        \label{fig:cf_mse}
    \end{subfigure}
    \hfill
    \begin{subfigure}[b]{0.48\textwidth}
        \includegraphics[width=\textwidth]{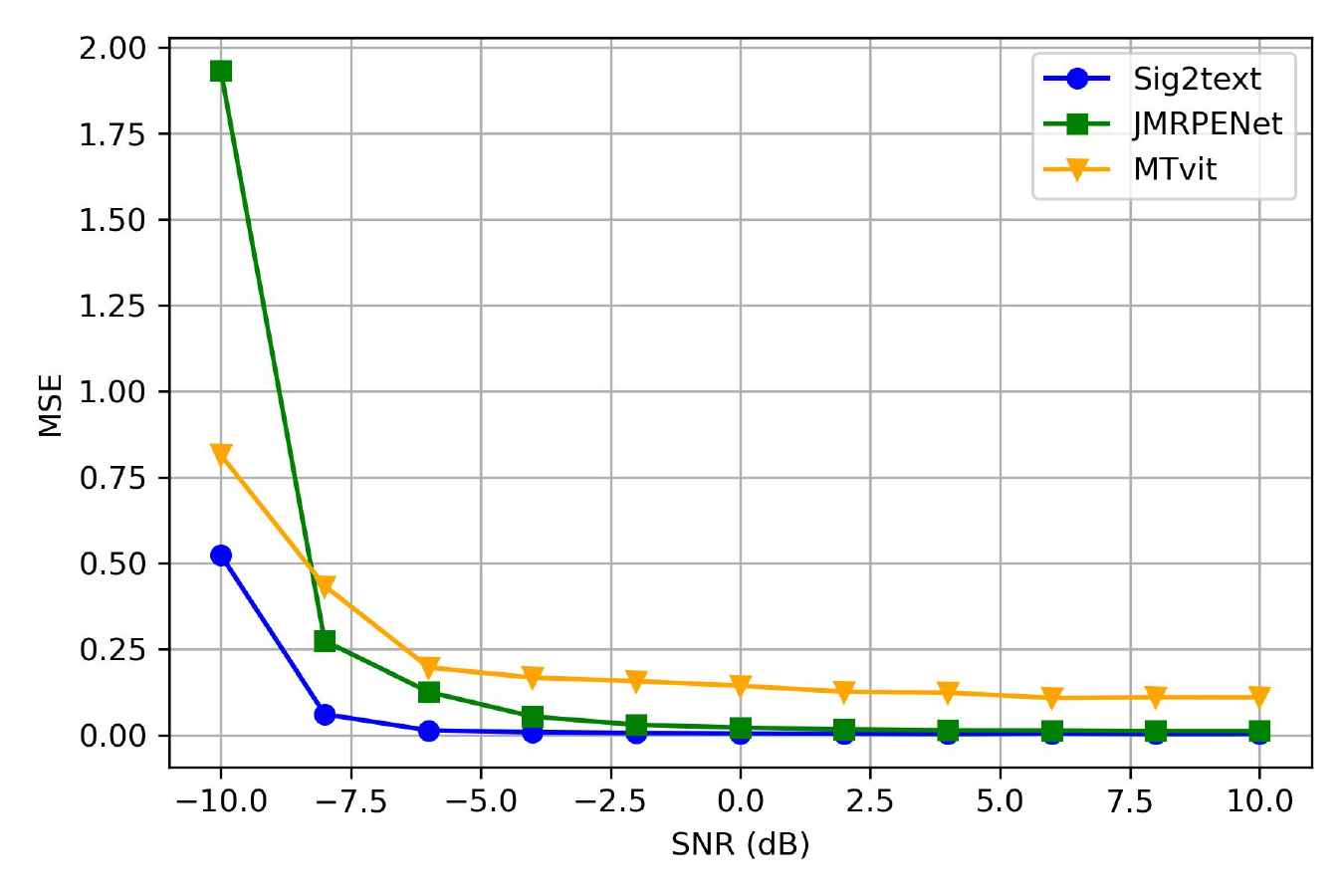}
        \caption{B}
        \label{fig:B_mse}
    \end{subfigure}
    \hfill
    \begin{subfigure}[b]{0.48\textwidth}
        \includegraphics[width=\textwidth]{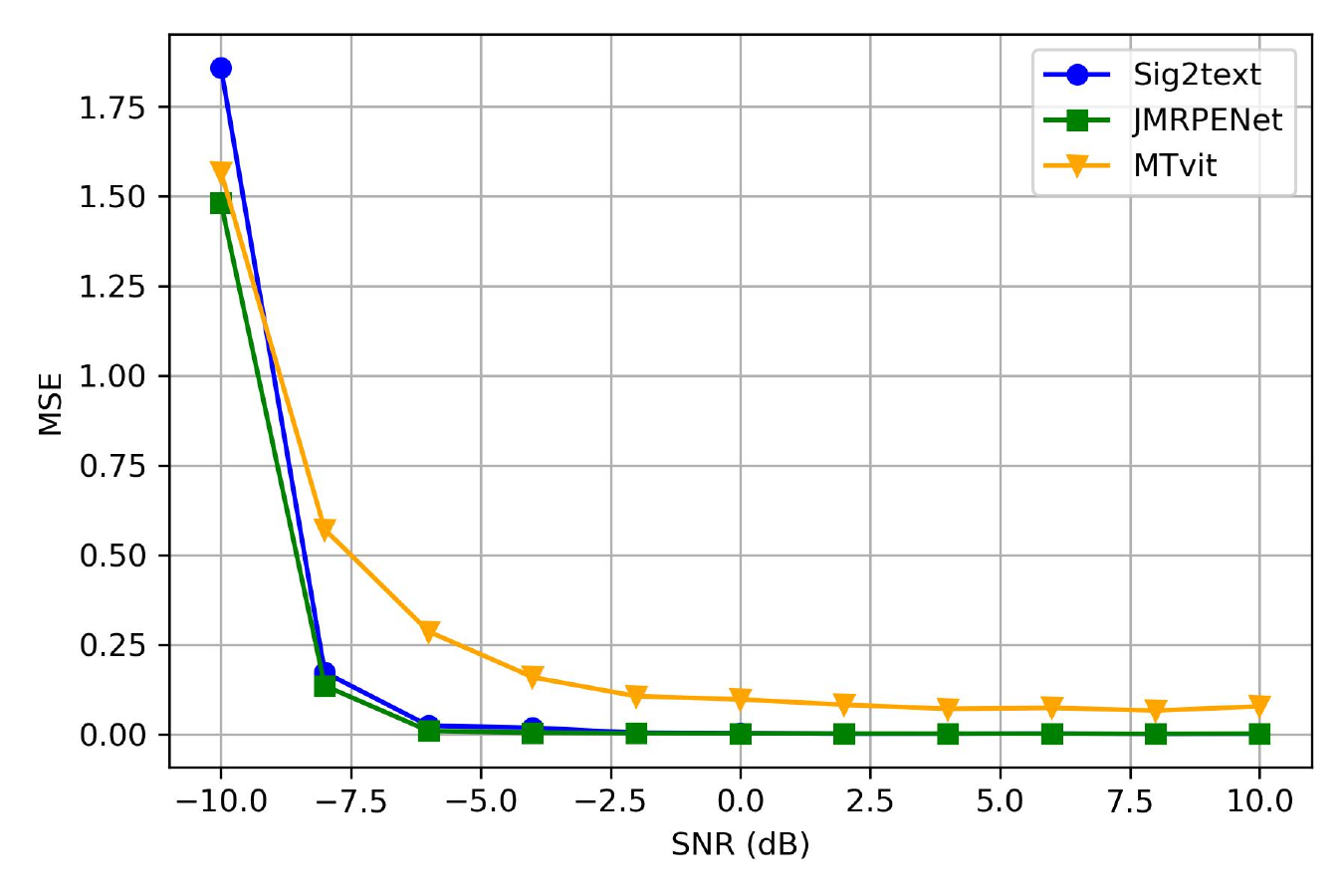}
        \caption{T}
        \label{fig:T_mse}
    \end{subfigure}
    \hfill
    \begin{subfigure}[b]{0.48\textwidth}
        \includegraphics[width=\textwidth]{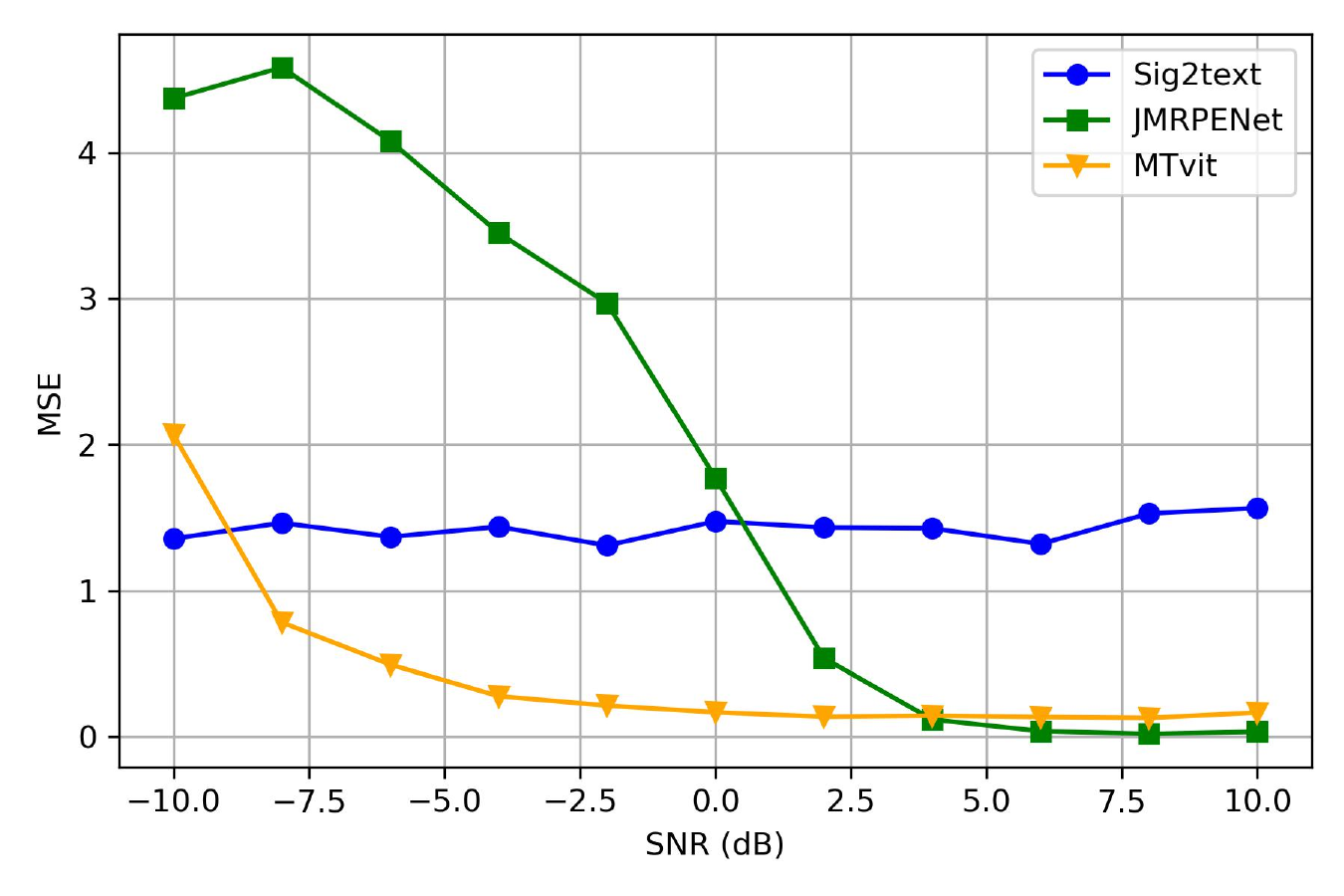}
        \caption{segment\_num}
        \label{fig:segment_num_mse}
    \end{subfigure}
    \caption{Parameter estimation performance across different SNR levels.}
    \label{fig:mse_comparison}
\end{figure}

\begin{figure}[htbp]
    \centering
    \begin{subfigure}[b]{0.48\textwidth}
        \includegraphics[width=\textwidth]{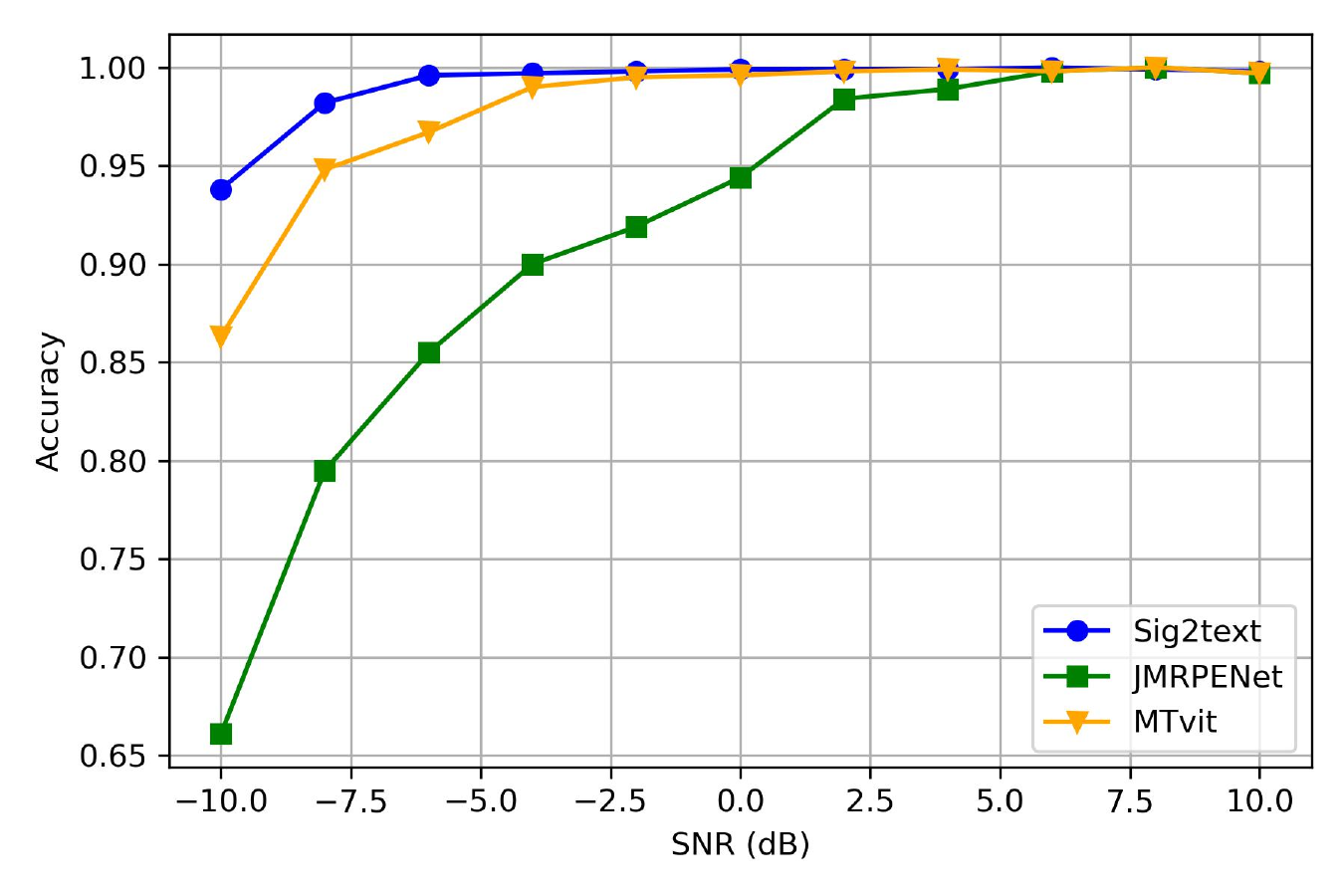}
        \caption{Type Recognition Accuracy}
        \label{fig:type_accuracy}
    \end{subfigure}
    \hfill
    \begin{subfigure}[b]{0.48\textwidth}
        \includegraphics[width=\textwidth]{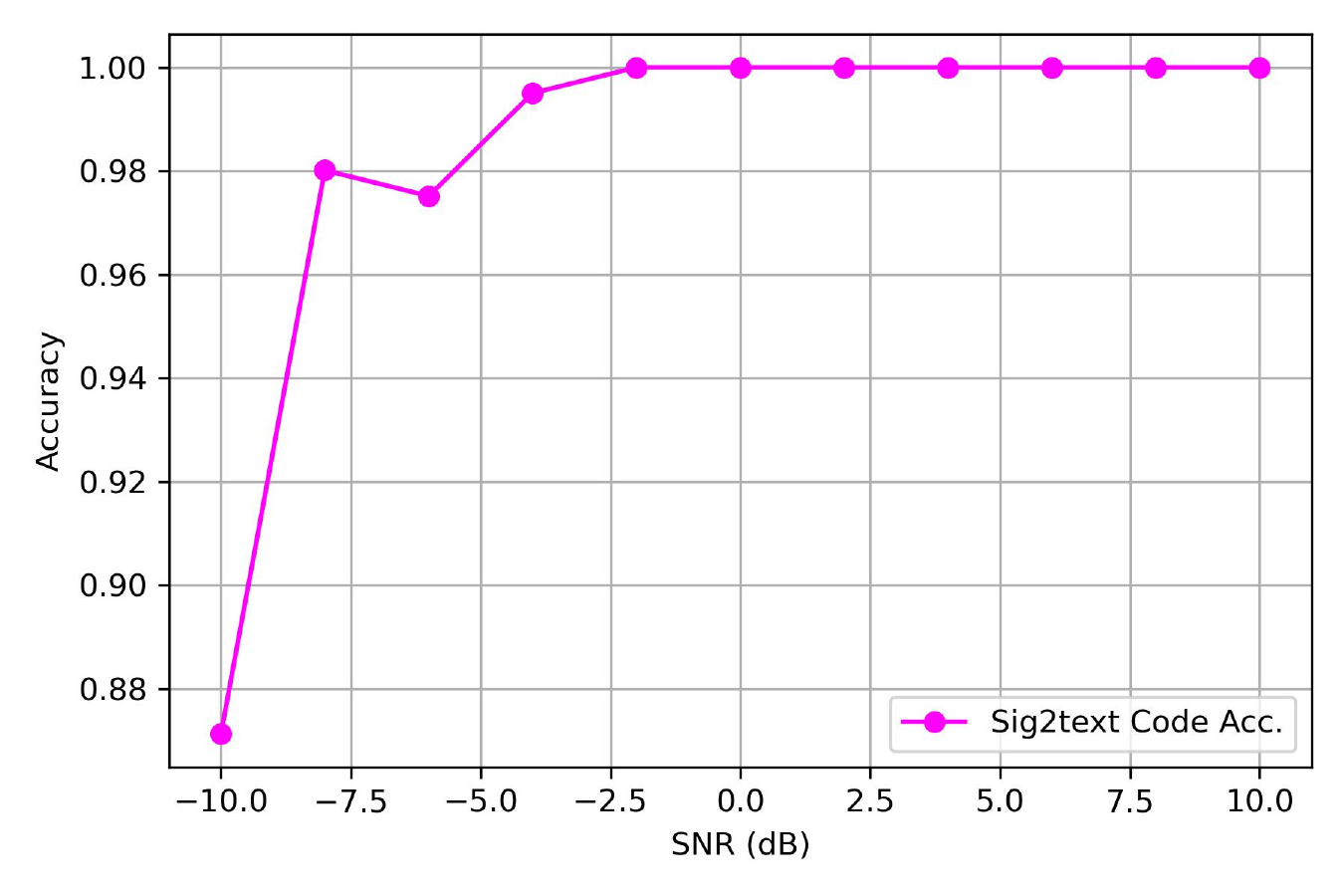}
        \caption{Code Estimation Accuracy}
        \label{fig:code_accuracy}
    \end{subfigure}
    \caption{Accuracy comparison and Sig2text code estimation performance.}
    \label{fig:accuracy}
\end{figure}

To further assess performance across varying noise conditions, we test the models at SNR levels ranging from $-10$ dB to $10$ dB. Figures~\ref{fig:mse_comparison} and~\ref{fig:accuracy} present the results for type recognition accuracy and the MSE of key parameters ($cf$, $B$, $T$, and $segment\_num$). Sig2text maintains superior parameter estimation performance for most metrics, though it struggles with segment number estimation. Notably, JMRPENet's performance improves with increasing SNR, supporting our initial hypothesis.

Figure~\ref{fig:type_accuracy} reveals that Sig2text achieves consistently higher type recognition accuracy across all SNR levels, maintaining over 94\% accuracy even at $-10$ dB. In contrast, JMRPENet shows strong SNR dependence, with accuracy dropping to 66\% at $-10$ dB but matching other models at 5 dB and above. This further validates the robustness of TFI-based features in noisy environments. Additionally, Sig2text demonstrates exceptional performance for Costas code sequences, achieving perfect estimation above $-2.5$ dB and maintaining reasonable robustness at lower SNR levels.

\begin{figure*}[htbp]
    \centering
    \begin{subfigure}[b]{0.32\textwidth}
        \includegraphics[width=\textwidth]{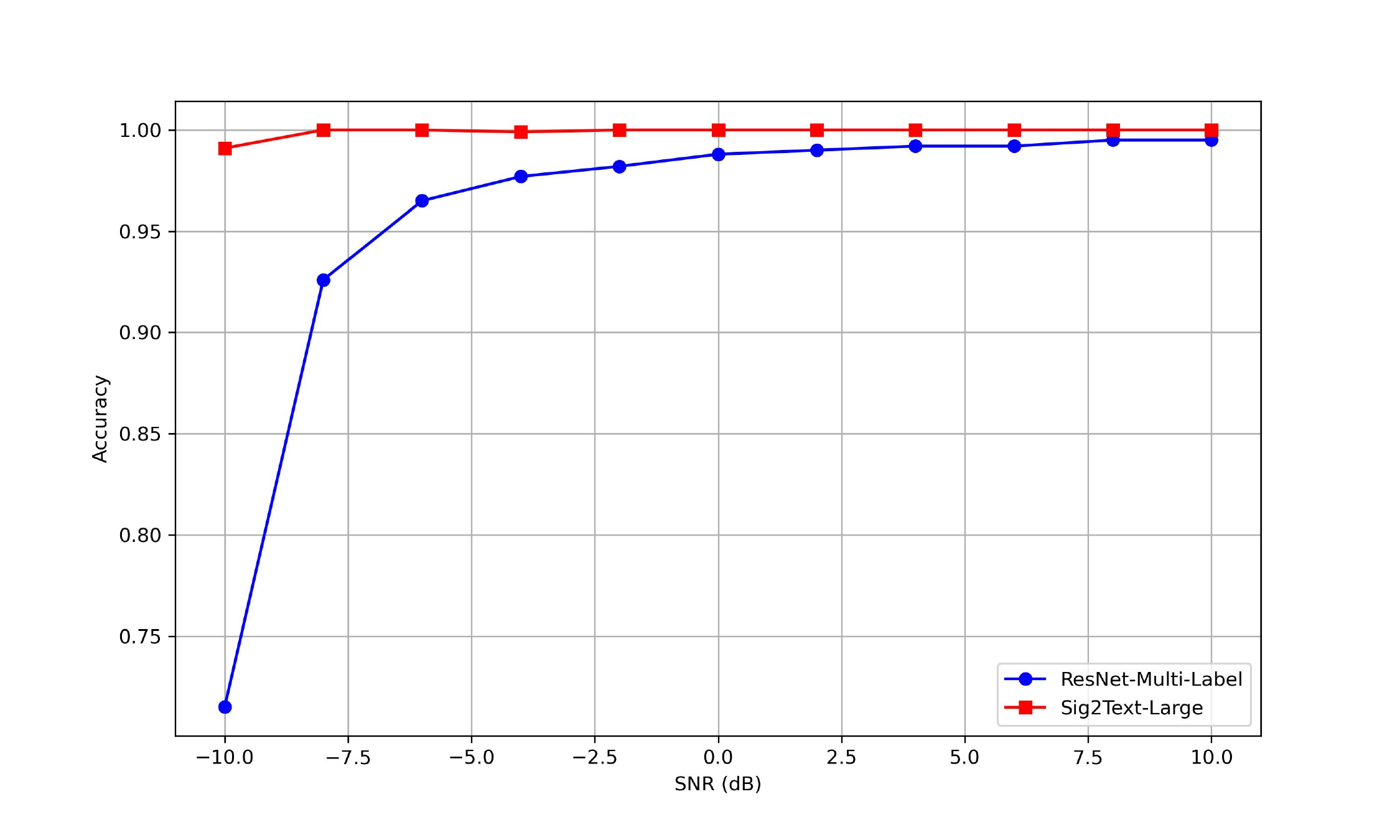}
        \caption{FM-FC Accuracy}
        \label{fig:acc_fm_fc}
    \end{subfigure}
    \begin{subfigure}[b]{0.32\textwidth}
        \includegraphics[width=\textwidth]{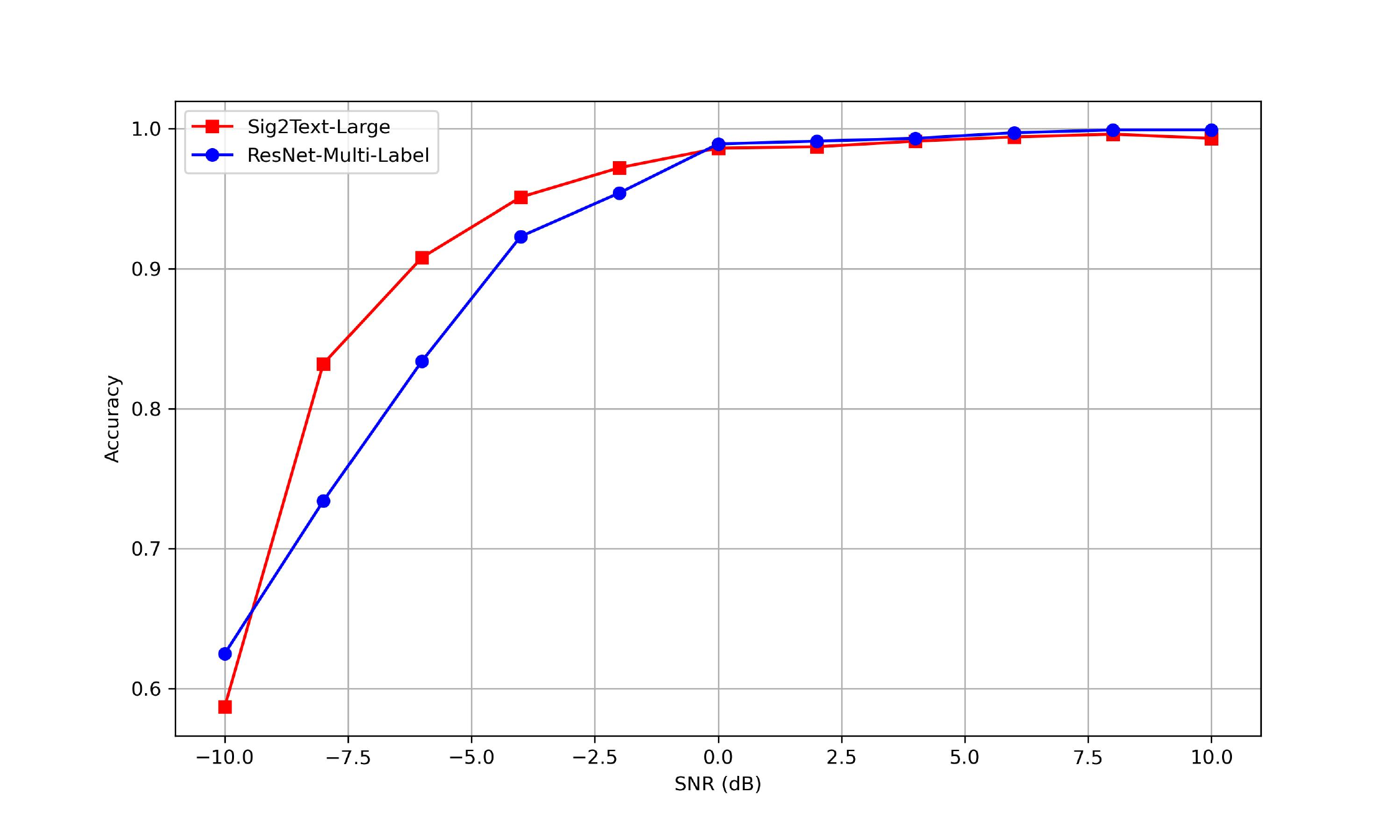}
        \caption{FM-PC Accuracy}
        \label{fig:acc_fm_pm}
    \end{subfigure}
    \begin{subfigure}[b]{0.32\textwidth}
        \includegraphics[width=\textwidth]{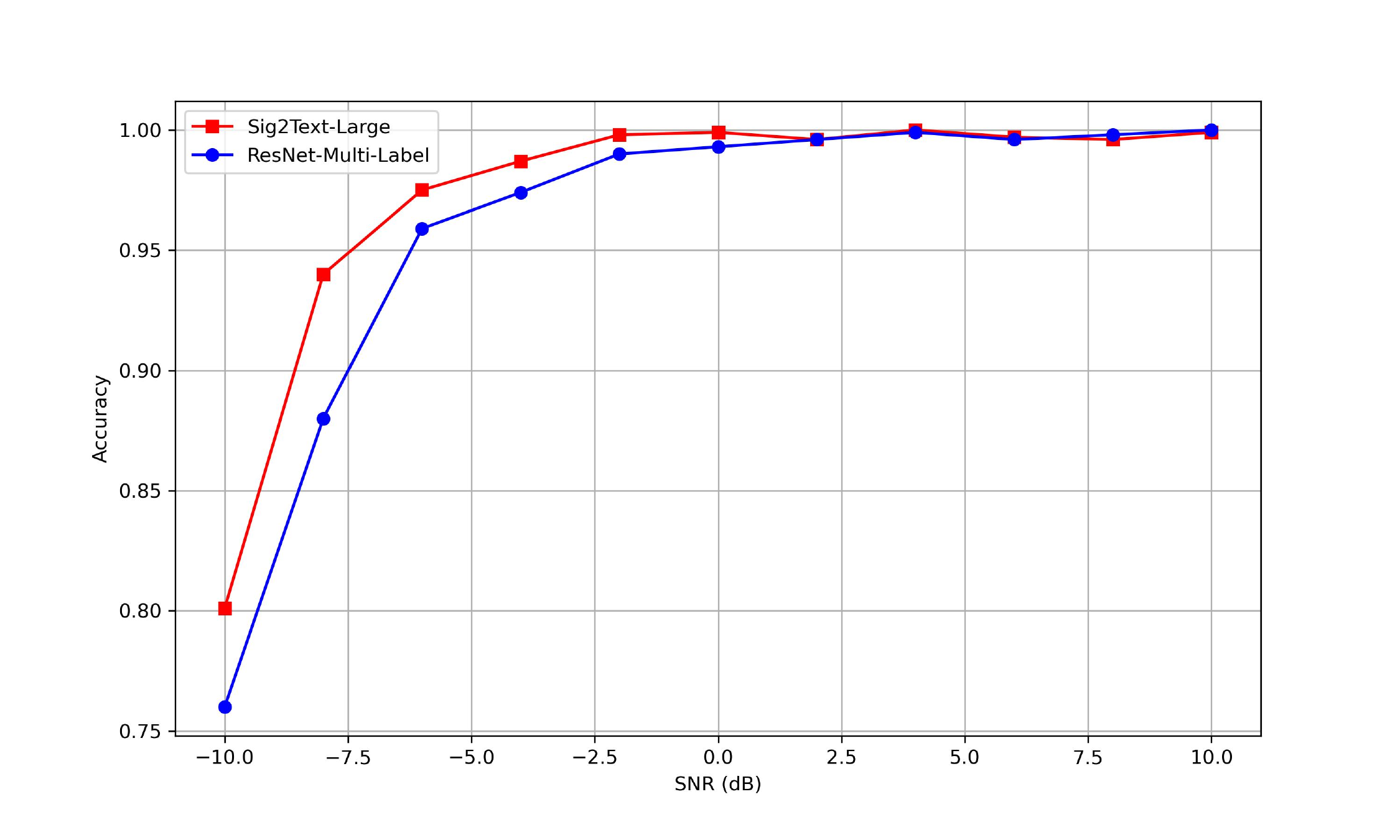}
        \caption{PC-FC Accuracy}
        \label{fig:acc_pm_fc}
    \end{subfigure}
    \begin{subfigure}[b]{0.48\textwidth}
        \includegraphics[width=\textwidth]{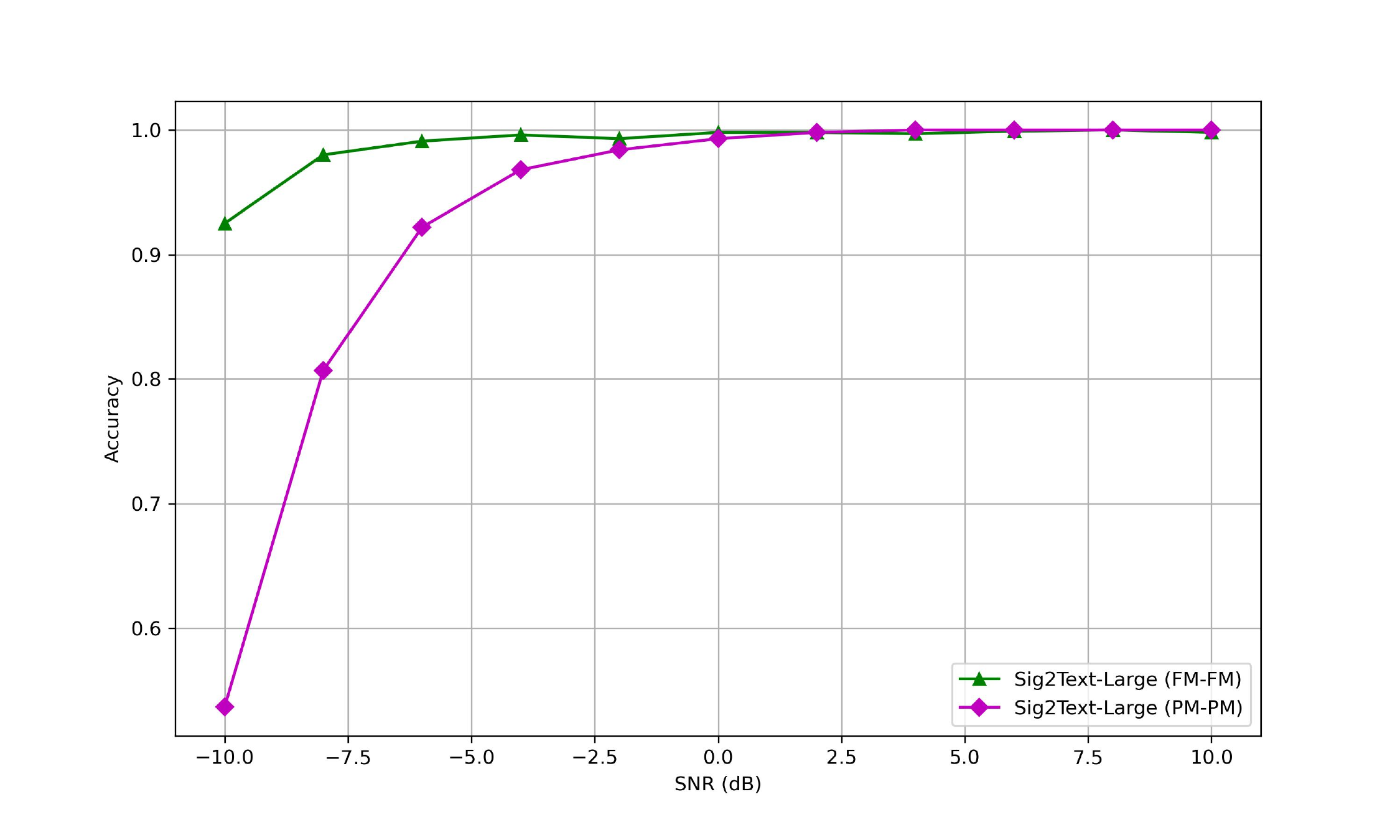}
        \caption{FM-FM and PM-PM Accuracy}
        \label{fig:acc_fm_fm_pm_pm}
    \end{subfigure}
    \begin{subfigure}[b]{0.45\textwidth}
        \includegraphics[width=\textwidth]{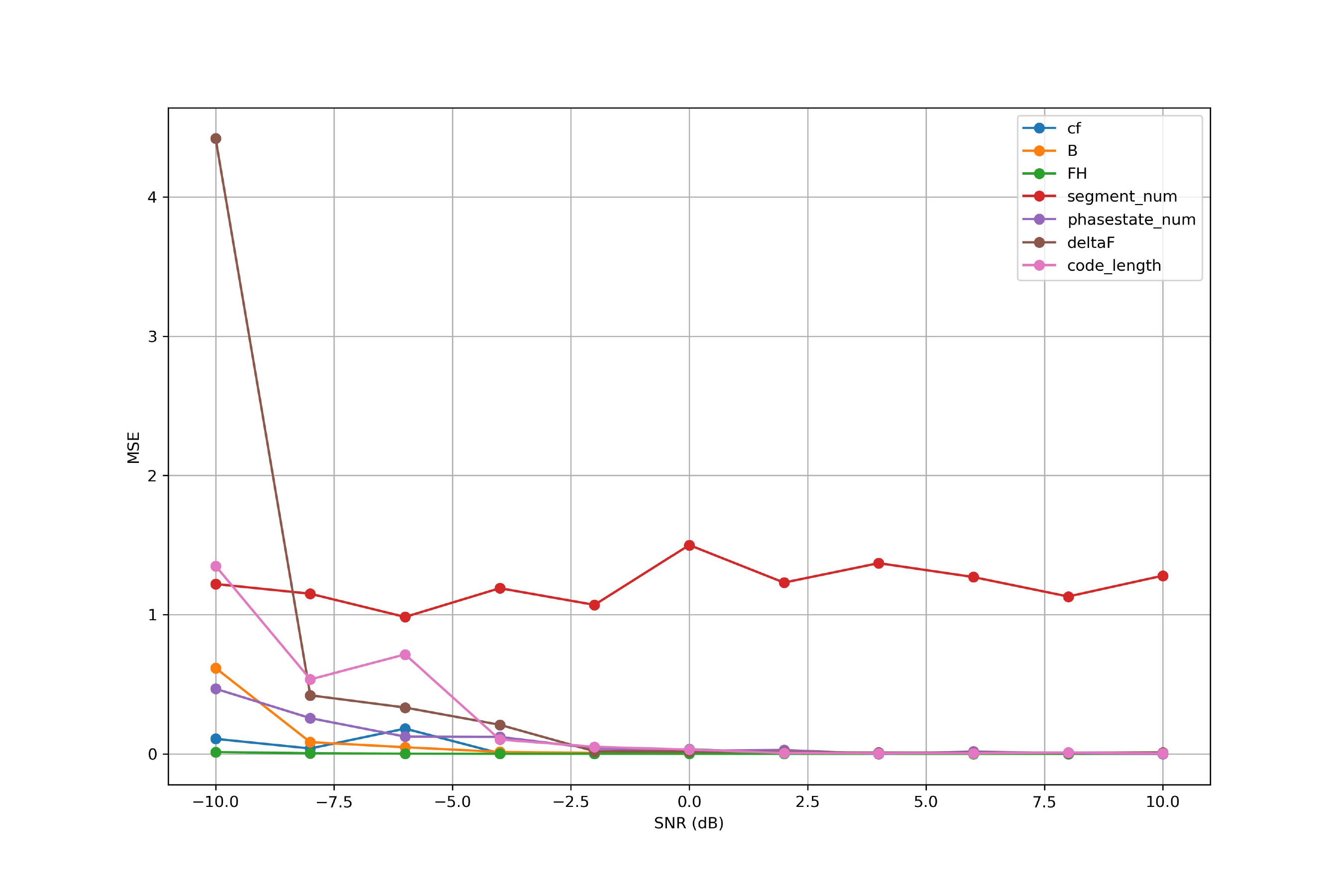}
        \caption{All Combinations - Parameters MSE}
        \label{fig:mse_all}
    \end{subfigure}
    \caption{Performance comparison between ResNet-Multi-Label and Sig2Text-Large. (a--c) Recognition accuracy for different signal combinations. (d) Accuracy of Sig2Text-Large for FM-FM and PM-PM combinations. (e) Parameter estimation MSE across all combinations.}
    \label{fig:results_comp}
\end{figure*}

Next, we evaluate the models on hybrid modulation signals. Using the same evaluation protocol, we report type recognition accuracy and parameter estimation MSE for different signal combinations (Fig.~\ref{fig:results_comp}). Sig2text-Large outperforms ResNet-Multi-Label across all combinations. Furthermore, it successfully processes same-type combinations while maintaining accurate parameter estimation.

\begin{figure}[htbp]
    \centering
    \includegraphics[width=0.8\linewidth]{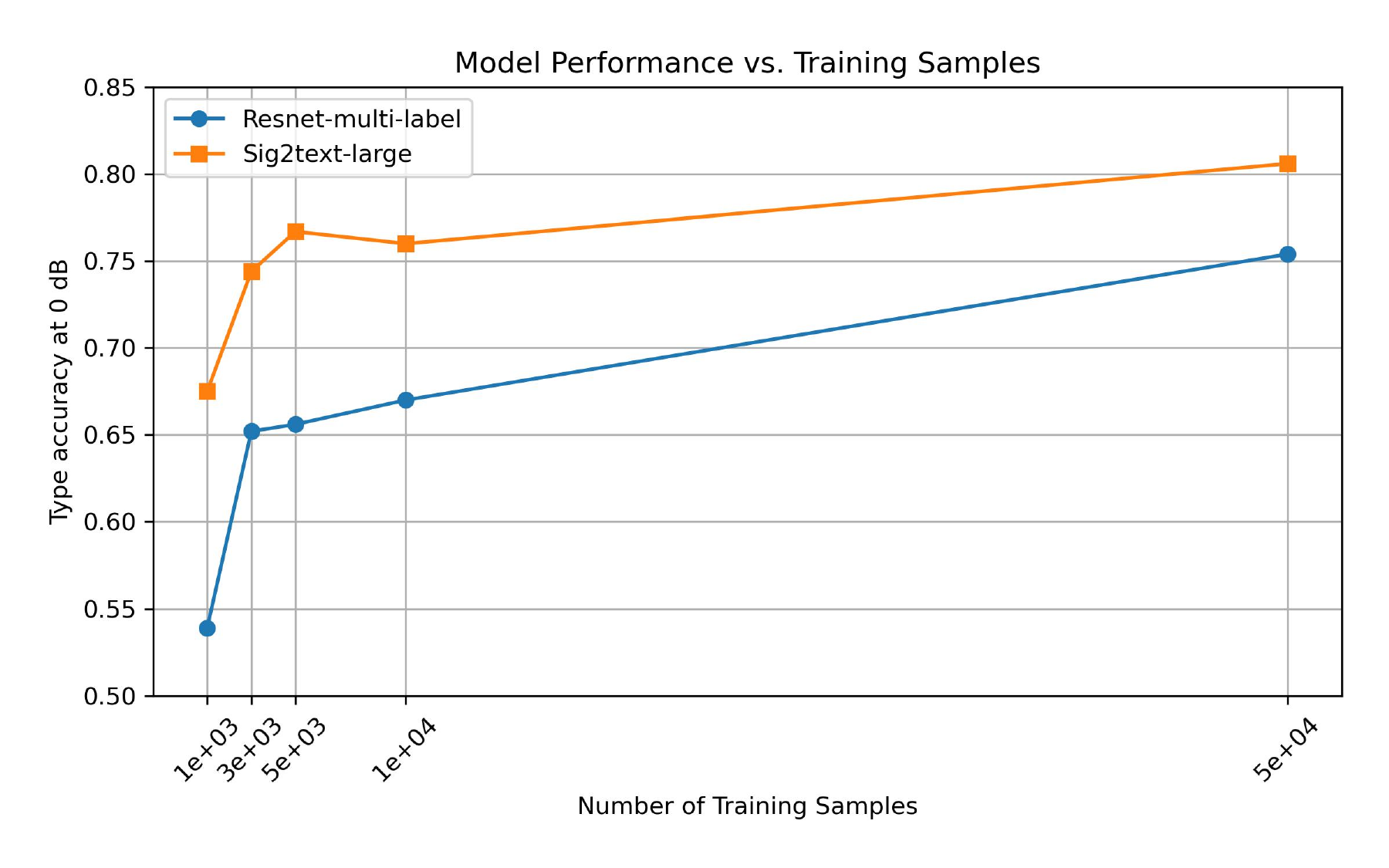}
    \caption{Type recognition accuracy after fine-tuning with different training sample sizes.}
    \label{fig:trans_performance}
\end{figure}

Finally, we examine the models' transfer learning capability by replacing linear frequency modulation (LFM) with sinusoidal modulation in Table~\ref{tab:modulation_parameters} and fine-tuning with varying amounts of training data (Fig.~\ref{fig:trans_performance}). Sig2text significantly outperforms multi-label ResNet, particularly with limited training samples, suggesting that its vision-language learning paradigm extracts more generalizable features.

We also measure inference speed by processing one sample of length 0.1ms on an NVIDIA A100 GPU with a batch size of 128 (Table~\ref{tab:model_summary}). Non-autoregressive models (MVit, ResNet-Multi-Label, JMRPE-Net) achieve real-time processing, while Sig2text is slower due to its larger size and autoregressive decoding. However, model compression techniques \cite{zhu2024survey} (e.g., quantization and pruning) could enable real-time operation.

\begin{table}[h]
    \centering
    \caption{Inference speed (including STFT computation) of different models}
    \label{tab:model_summary}
    \begin{tabular}{lcc}
    \toprule
    \textbf{Model}          & \textbf{Inference Speed} & \textbf{Model Inputs}          \\
    \midrule
    Sig2text               & 0.8 ms                  & STFT magnitude patches         \\
    Sig2text-Large         & 4 ms                    & STFT magnitude patches         \\
    MVit                   & 0.020 ms                & STFT magnitude patches         \\
    ResNet-Multi-Label     & 0.045 ms                & STFT magnitude                 \\
    JMRPE-Net              & 0.015 ms                & Raw waveform                   \\
    \bottomrule
    \end{tabular}
\end{table}
\section{Conclusion}
In this paper, we propose Sig2text, a unified approach for radar signal recognition and parameter estimation. The method employs a Vision Transformer architecture to extract time-frequency features from radar signals, coupled with a transformer-based decoder that parses symbolic representations defined by a CFG. We evaluate the proposed method on a synthetic radar signal dataset and compare its performance against baseline approaches. Experimental results demonstrate that Sig2text achieves superior performance in both signal type recognition and parameter estimation tasks. Furthermore, the model exhibits strong transfer learning capabilities, suggesting its ability to learn generalizable feature representations.

Several promising directions exist for future research. First, the method could be extended to perform signal detection tasks by augmenting the CFG productions with additional tokens representing pulse boundaries (e.g., \texttt{<toa>} and \texttt{<toe>}). Second, similar to the representation of hybrid modulations, the CFG could be further expanded to accommodate overlapping signals, thereby enabling the model to handle more complex real-world scenarios.

\bibliographystyle{plain} 
\bibliography{LLM_radar_paper.bib} 
\end{document}